%% file: mian.tex
\documentclass{article}
\usepackage{ijcai19}

\usepackage{times}
\usepackage{soul}
\usepackage{url}
\usepackage[hidelinks]{hyperref}
\usepackage[utf8]{inputenc}
\usepackage[small]{caption}
\usepackage{graphicx}
\usepackage{amsmath}
\usepackage{booktabs}
\usepackage{algorithm}
\usepackage{algorithmic}
\usepackage{color}
\usepackage{todonotes}
\definecolor{notSureColor}{RGB}{147, 9, 229}
\colorlet{ColorforSina}{green!5!orange!95!}

\usepackage{enumitem}
\usepackage{multirow}% http://ctan.org/pkg/multirow
\usepackage{hhline}% http://ctan.org/pkg/hhline

\usepackage{makecell}

\usepackage{bbding}
\usepackage{pifont}
\usepackage{wasysym}
\usepackage{amssymb}

\title{Open Issues in Combating Fake News: Interpretability as an Opportunity}
\author{
Sina Mohseni$^1$
\and
Eric D. Ragan$^2$\and
Xia Hu$^{1}$ % \And
\affiliations
$^1$Texas A\&M University, 
$^2$University of Florida\\
\emails
\{sina.mohseni, xiahu\}@tamu.edu,
eragan@ufl.edu
}
\begin{document}
\maketitle

\input{files/0-abstract.tex}
\input{files/1-intro.tex}

\input{files/2.1.life-cycle.tex}

% \input{files/2.2.method-review.tex}
% \input{files/2.3.interpret-dimensions.tex}
\input{files/3-problem.tex}

\input{files/3.1.clarity.tex}

\input{files/3.2.newsfeed.tex}
\input{files/4-discussion.tex}

\input{files/4.1.human-interpret.tex}

\input{files/4.2.transparency.tex}
\input{files/5-conclusion.tex}

\bibliographystyle{named}
\bibliography{Bibliography}

\end{document}

%% file: files/0-abstract.tex
\begin{abstract}
Combating fake news needs a variety of defense methods.
Although rumor detection and various linguistic analysis techniques are common methods to detect false content in social media, there are other feasible mitigation approaches that could be explored in the machine learning community. 
In this paper, we present open issues and opportunities in fake news research that need further attention. 
We first review different stages of the news life cycle in social media and discuss core vulnerability issues for news feed algorithms in propagating fake news content with three examples.
We then discuss how complexity and unclarity of the fake news problem limits the advancements in this field. 
Lastly, we present research opportunities from interpretable machine learning to mitigate fake news problems with 1) interpretable fake news detection and 2) transparent news feed algorithms.
We propose three dimensions of interpretability consisting of algorithmic interpretability, human interpretability, and the inclusion of supporting evidence that can benefit fake news mitigation methods in different ways.
\end{abstract}

%% file: files/1-intro.tex
\section{Introduction}

% \Title{Problem 1: Fake news in Social media}
Fake news has many faces. 
Allcott and Gentzkow~\shortcite{allcott2017social} define fake news as an article that is intentionally created and verifiably false, however, it may hide behind deceptive writing or behind an innocent headline.
% It may look innocent and circulate like a rumor, but still be falsified information. % from a credible source 
In practice, every time we close a door to prevent the propagation of fake news, it enters from another. 
Fake news does not only spread due to intentional and malicious dissemination, but also thrives simply because it is an easy sell and commonly has a very large audience through social media.
Although social media was primarily designed to benefit people with information sharing, nowadays it also contains many forms of misinformation and disinformation.
% Finding a variety of fake new related phenomena needs diving deeper into different types and purpose of inaccurate information from other disciplines.  
% For example, while satire news could be an entertaining type of disinformation and click baits could be written for advertisements, the fake news is there to do harm.
The complex nature of news veracity analysis and multi-modal information sources contribute to the fake news identification problem.
Even though many fake news related concepts have been well studied in machine learning research, the grand challenge of fake news is yet to be overcome.
The importance of analyzing information truthfulness is greater than ever with the increasing popularity and impact of social media among people.
 
% \Title{Problem 2: Data misuse in social media}
Besides fake news detection methods as a primary solution, news-feed algorithms (identical to news recommenders in most cases) are also vulnerable to data misuse and attacks that could result in the spread of fake news.
The popularity of social media has resulted in large, continuously updating collections of user data.
Alongside users' activity in sharing information to socialize and express their opinions in a virtual social environment, machine learning algorithms constantly process user activity and content for personalized content and targeted advertisement on large scale. 
However, with the growing amount of user data in social media, the implications of personalized data for the dissemination and consumption of news has caught the attention of many, especially given evidence of the influence of malicious social media accounts on the spread of fake news to bias users during the 2016 US election~\cite{bessi2016social}.
Studies show the targeted distribution of erroneous or misleading ``fake news'' may have resulted in large-scale manipulation of users' news feeds as part of the intense competition for attention in the digital media space~\cite{Social2018Inequalities}.

% \Title{Problem 3: Interpretability and social media}
Interpretable machine learning algorithms are gaining increasing attention as people are interacting more with machine learning products in day-to-day life~\cite{gunning2017explainable}.
Interpretable algorithms in social media applications can assist users in identifying biased algorithms by explaining why a certain recommendation or decision is made.
Transparency is an essential need since biased algorithms at the decision-making and information-distribution levels may cause unintentional discrimination that could result in the loss of opportunities or social stigmatization at a large scale.
Lack of explanation of how the content is selected for the user may result in unaware users who think they have access to all available information rather than only a small subset of news.
For example, research shows personalized news feed algorithms are not immune to bias~\cite{bozdag2013bias} and can even cause intellectual isolation~\cite{pariser2011filter} over time.

In this paper, we present open issues and opportunities in fake news research that would benefit from further attention from the machine learning community. 
Rather than focusing only on fake news detection methods, we also present research opportunities to mitigate the circulation of false content in social media. 
We first review current fake news detection and mitigation methods at different stages of the news life cycle: \textit{creation} of news, \textit{distribution} of news, and \textit{consumption} of news.
Next, we open the discussion on open issues in fake news research. 
We review how complexity and unclarity of the fake news detection problem limit the advancements in this field.
For a broader investigation, we also discuss the vulnerability of news feed algorithms in propagating fake news content, affecting news diversity, and hindering credibility.
At last, we present interpretability as a solution for improving fake news methods through 1) interpretable fake news detection and 2) transparent news feed algorithms.
We propose three dimensions of interpretability consisting of algorithmic interpretability, human interpretability, and the inclusion of supporting evidence that can benefit fake news mitigation methods in different ways.

%% file: files/2.1.life-cycle.tex
\begin{figure}[t!]
\centering
  \includegraphics[width=0.99\columnwidth]{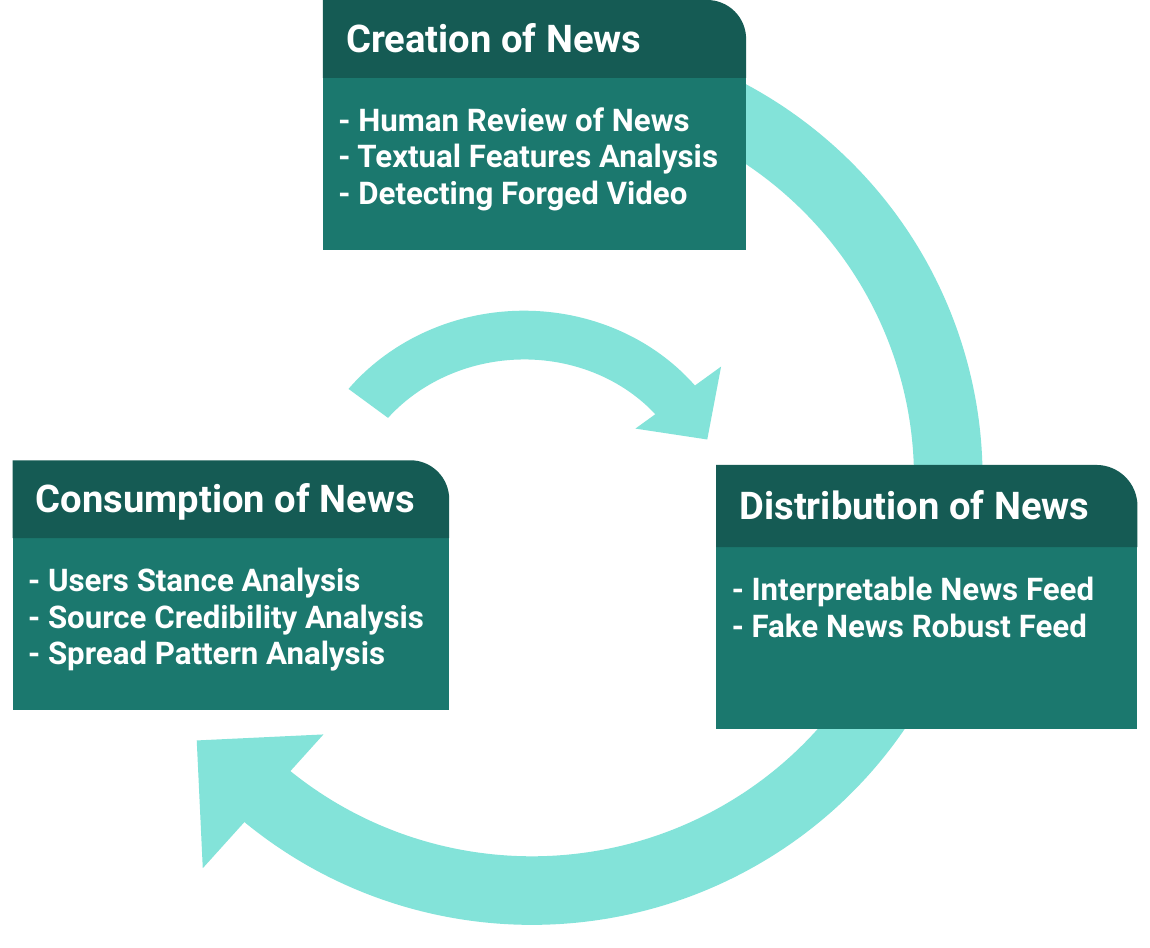}
  \caption{A summary of fake content detection methods at different stages of the news life in social media.
  While natural language processing and social media data mining methods are popular in fake news at creation and consumption stages, there is a limited amount of research on fake news robust news feed algorithms to mitigate the propagation of fake content.
%   Although interpretable news feed is not a detection or debunking solution, in real scenarios, mitigation solutions could be more advantageous than automatically debunking of the news
%   to mitigate the propagation of fake content.
  The inner arrow shows how user's social data is used to personalize news recommendation algorithms.
  }
  \label{fig:life-cycle}
\end{figure}

% Link: https://docs.google.com/presentation/d/1Jbhuy4gYMVbmvcyJe0-R7Fr7V9ncFu4KaYntNxbHN70/edit#slide=id.g456c348b22_0_77

% \checkmark
% \Checkmark
% \CheckmarkBold
% \ding{51}
% \ding{52}
% \CheckedBox

\begin{table*}
    \centering
  \caption{2D characterization of inaccurate information types and their detection methods. Different data and approaches are employed to detect various misinformation and disinformation types, however, identifying fake news involves knowledge based fact checking of statements.}
  \label{tab:main-table}
\begin{tabular}[t] {|c|c||c|c|c|c|c|c|c|c|c|}%{ |p{1.4cm}||p{1.4cm}|p{1.4cm}|p{1.4cm}|  }
 \hline
%  \multicolumn{4}{|c|}{Country List} \\
%  \hline
%  \multirow{3}{4em}{Multiple row} &
  \multicolumn{2}{|c||}{\textbf{Analysis Types}} &
  \multicolumn{9}{c|}{\textbf{Inaccurate Information Types}} \\
 \hhline{~~---------}
  \multicolumn{2}{|c||}{ } &
  \multicolumn{2}{c|}{\textbf{Misinformation}} & \multicolumn{7}{c|}{\textbf{Disinformation}} \\
  \hline 
  Approach & Methods & \makecell{False Info.} & \multicolumn{2}{c|}{ Rumors} & Click bait & \makecell{Satire \\News} & \makecell{Fake \\News} & \makecell{Deceptive \\ News} &  Spam & \makecell{Fake\\ Review} \\
  \hline
  \makecell{Knowledge \\ Based} & \makecell{Knowledge Graph \\ Fact Checking} & \makecell{ \ding{51} \\  \ding{51}}  &  \multicolumn{2}{c|}{ \makecell{-\\-} } & \makecell{- \\ -} & \makecell{-\\ -} & \makecell{ \ding{51} \\  \ding{51}}  & \makecell{-\\ -} & \makecell{-\\ -} & \makecell{-\\ -} \\
  % & Y & & & \\
  \hline
  \makecell{Style \\ Based} & \makecell{Headline Analysis \\ Deceptive Analysis} & \makecell{- \\ -}  & \multicolumn{2}{c|}{ \makecell{- \\ -} } & \makecell{\ding{51} \\ -} & \makecell{- \\ \ding{51}} & \makecell{ - \\ - }  & \makecell{- \\ \ding{51}} & \makecell{\ding{51} \\ \ding{51}}& \makecell{-}\\
  \hline
  \makecell{Social \\ Context \\ Based} & \makecell{Source Credibility \\ Comment Credibility \\ Propagation Pattern \\ Network Analysis} & \makecell{\ding{51} \\- \\ -\\ -}  & \multicolumn{2}{c|}{ \makecell{- \\- \\ \ding{51}\\ \ding{51}}} & \makecell{- \\- \\ -\\ -} & \makecell{- \\- \\ - \\ -} & \makecell{- \\ - \\ - \\ -}  & \makecell{- \\ - \\ -\\ -}  & \makecell{\ding{51} \\- \\ -\\ -}& \makecell{-\\ \ding{51} \\ -\\ -}\\  % spam - deceptive news  
%   \hline
%   \makecell{Propagation \\ based} & \makecell{- \\ -} & \makecell{- \\ -}  &  \multicolumn{2}{c|}{ \makecell{- \\ -} } & \makecell{- \\ -} & \makecell{- \\ -} & \makecell{- \\ -} \\
%   Style based    &  &  &  &  &  \\
%   Social Based   &  &  &  &  &  \\
%   Propagation Based  &  &  &  &  &  \\
 \hline
\end{tabular}

\end{table*}

\section{News Life Cycle in Social Media}
% \section{Analyzing Fake Content in the News} 
% --Mention early stages of fake news -- 

Different fake news surveys (e.g.,~\cite{shu2017fake},~\cite{zhou2018fake},~\cite{sharma2019combating} and~\cite{zubiaga2018detection}) provide comprehensive reviews of fake news definitions, data mining methods, training data sets, and recognition metrics for fake news research.
However, these works lack to acknowledge an important missing piece in the current fake news mitigation research.
In this section, we emphasize on a research gap in studying effects of news recommendation algorithms on the spread of fake content and crediting unreliable sources (see Figure~\ref{fig:life-cycle}). 
In order to open the discussion toward research limitations and opportunities, we briefly review current fake content detection approaches during three main stages of news life cycle in social media.
% Figure~\ref{fig:chain} shows news life cycle in social media and potential solutions to mitigate fake news at different stages.

\subsection{Creation of News}
The first stage is to detect fake content at the news creation step which traditionally is done with human review through expert review and crowdsourcing techniques at the early stages.
Experts review the truthfulness of the news by evidence and determine whether claims are accurate or false (partially or entirely).
Fact checking is a knowledge-based approach usually done by fact-checking organization (e.g., Politifact.com and Snopes.com) to judging the veracity of news piece with external references~\cite{vlachos2014fact}. %\footnote{http://www.politifact.com} \footnote{https://www.snopes.com}
However, expert-review fact-checking methods are time-consuming, expensive, and not scalable for stopping the spread of fake content in social media. 

Machine learning methods also analyze falsified context (not limited to news or social media) using various types of data mining and machine learning techniques.
One approach, for instance, is to use linguistic features to analyze writing styles to detect possible false content~\cite{afroz2012detecting}.
For example, recognizing deception-oriented~\cite{rubin2015truth} and hyper-partisan content~\cite{potthast2017stylometric} can be used as a basis for detecting intentionally falsified information.
Also, various spam detection \cite{spirin2012survey} and satire news detection~\cite{rubin2016fake} methods.
Another approach is to use clickbait detection algorithms~\cite{potthast2016clickbait} to analyze inconsistency between headlines and content of the news for possible fake news detection.
% \TODO{describe our work} ~\cite{afroz2012detecting} ~\cite{yang2019XFake}

% other data type 
Additionally, fact-checking is not limited to the correctness of textual content---images and videos may be evaluated or used as evidence as well. 
For the cases of forged images and videos, researchers use deep learning methods (e.g.,~\cite{julliand2015image,afchar2018mesonet}) to detect falsified contents.
New methods like provenance analysis have also been utilized for content validation via generating provenance graph of images as the same content is shared and modified over time~\cite{bharati2018beyond}.

\subsection{Distribution of News}
The next stage of news life in social media is the distribution of content via search engines and news feed algorithms. 
Although distribution of news can be an effective stage to combat fake news distribution, machine learning research communities paid little attention to the importance of news search engines, news recommender algorithms, and daily news feeds in propagating fake news.
% Others also examined the use of spacial news feed in virtual reality environment as opposed to linear feeds to reduce the creation of filter bubbles~\cite{linder2018pop}.
For example, echo chambers (discusses in section~\ref{echo-chamber}) are from social media vulnerability points that create and propagate false information. 
Multiple sources of evidence show personalized news feed algorithms can have drastic effects on news diversity and the creation of echo chambers in social media.
In a recent study by Geschke et al.~\shortcite{geschke2018triple} presented an agent-based simulation of different information filtering scenarios that may contribute to social fragmentation of users into distinct echo chambers.
Their observation of agent-based modeling found that social media filters can boost social polarization and lessen the interconnections of social media echo chambers.
In the next sections, we further discuss the accountability of news feed algorithms for propagating fake content by creating echo chambers and propose interpretability as a potential solution for this problem.

% ---------- diversity metrics ---------------

% Researchers also propose using diversity metrics~\cite{valcarce2018robustness,ekstrand2018all} and bias quantification methods~\cite{kulshrestha2017quantifying} as other ways to study bias and discrimination in recommendation algorithms.
% For example, Kulshrestha et al.~\shortcite{kulshrestha2017quantifying} proposed a framework to quantify bias in ranked search results in political-related queries on Twitter.
% Their framework can distinguish bias from news content and ranking algorithm, and they found evidence of significant effects of both input content and search algorithms in producing bias.

% Despite the research on detecting the falsified content by its textual and social features, current state of machine learning research is short on studying the critical role of news recommendation algorithms in distribution of false content and their vulnerability for being misused and causing discrimination. % \cite{ma2019detect}

% The importance of news feed algorithms is their critical role in providing news content and their vulnerability for being misused and causing discrimination.
% In reviewing open issues in fake news research, we further discuss the accountability of news feed algorithms for propagating fake content by creating echo chambers and filter bubbles in social media and news search engines.
% We also suggest interpretable news feed algorithm will solve these problems by increasing user awareness, holding users accountable for their content, and exposing users to opposite political views.

\subsection{Consumption of News}
The final stage of analyzing fake news is to process social media users' stance, analyze news propagation patterns, and estimate news source credibility to find possible false content in social media.
Research on social media data mining show dominant results in detecting fake content and account, however, utilizing social media data means waiting until the fake content is already exposed to the users.
This shows a trade-off between leveraging rich social data for fake news detection and waiting until a group of users is exposed to fake content.

% -- machine learning --
In addition to analyzing the news by their content, other social media information such as source credibility~\cite{castillo2013predicting}, users' stance~\cite{jin2016news}, and news temporal spreading pattern~\cite{kwon2013prominent} have been used to assess the veracity of the news.
Such social features can be applied to user groups to evaluate the credibility of specific news pieces by considering the stance of a group of users for the news topics~\cite{tacchini2017some}.
Similarly, rumor detection methods aim to detect a track of posts discussing a specific topic~\cite{ma2016detecting}.

To increase fake news detection accuracy and model generalizability, training on multi-source and multi-modal datasets are also studied.
For example, Shu et al.~\shortcite{shu2017exploiting} explored the correlation between news publisher bias, user stance, and user engagement together in their Tri-Relationship fake news detection framework.
In a following work, Shu et al.~\shortcite{shu2018fakenewsnet} proposed a training dataset to include news content and social context along with dynamic information of news. 

Although most aforementioned data mining methods do not perform a direct fake news detection, these methods can leverage both social and textual feature to identify suspicious news pieces for human review.

%% file: files/3-problem.tex
% \section{Limitations in Current Methods}
% \section{What holds Fake News Research Back}
% \section{What Slowdowns Fake News Research }
% \section{What are Fake News Mitigation Limits}
% \section{Limitations in Fake News Research}
\section{Open Issues in Fake News Research}
In the previous sections, we discussed current methods to detect and mitigate fake content at different stages of news life in social media.
Now we review two main issues in the fake news research that put a limit on the current state of the art systems.
% We also discuss how news feed algorithms and personalized search engines contribute to spreading fake content.

%% file: files/3.1.clarity.tex
\subsection{The (Un)clarity in Problem Characterization}

Problem characterization is the very first step in problem solving process.
Reviewing literature from multiple disciplines such a social science, psychology and machine learning shows various definitions for fake news and its related phenomena.
Although each represents a type of inaccurate information and news (i.e., information from current events), one can find at least seven different concepts related to the falsified information and news including hoax, fake news, rumors, deceptive news, spams, click baits, forged images, and videos, etc.
Although different in shape and purpose, these concepts share the same nature of inaccurate information: either in form of misinformation or disinformation.
In this section, we briefly review different fake news related phenomena and their relationship to the fake news problem. % , current machine learning solutions, and their limitations in real scenarios.

\subsubsection{Incomplete Problems}

Reviewing fake news related literature indicates a misinterpretation between false news detection and fake news recognition methods.
% Below we describe the difference between each of these tasks:
% \subsubsection{False News Detection}
Although machine learning researchers actively study different methods to detect the existence and circulation of various types of misinformation and disinformation in social media, yet the fake news gets a narrow definition today.
Allcott and Gentzkow~\shortcite{allcott2017social} define fake news as an article that is intentionally created and verifiably false. 
Here we review examples of misinformation and disinformation and compare their differences with fake news.
Various types of misinformation in forms of inaccurate posts (e.g., inaccurate scientific facts), rumor (e.g., inaccurate reporting of an event) and news (e.g., political and economic news) represent incomplete statements and events.
Also, disinformation is purposely created for financial and political gain (e.g., fake news and rumors), advertisement (e.g., click baits), entertainment (e.g., satire news and online memes), fame (e.g., forged photos and videos) and many other purposes.
Therefore in many cases, fake news could be a subcategory of disinformation and detecting fake news is not always equal to detecting any rumor or deceptive report. 

However,, in reality, fake news may still spread with rumor-like patterns in social media or may be embedded in a click-bait political advertisement.
This means, although various detection techniques could be effective in detecting the fake news (as well as other types of inaccurate information), this is not necessarily equal to fake news in its narrow definition.
The fake news problem clarification tells us detecting fake-new-related concept -- although might seem the similar -- is not the same as detecting fake news.

\subsubsection{Incomplete Solutions}

After meta-reviewing recent fake news papers and  related surveys (e.g.,~\cite{shu2017fake},~\cite{zhou2018fake},~\cite{sharma2019combating}, ~\cite{spirin2012survey} and~\cite{zubiaga2018detection}) to analyze different perspectives of fake news solutions, we present a 2D categorization of different misinformation and disinformation types and current machine learning detection methods in Table~\ref{tab:main-table}. 
This table shows although an ultimate solution for fake news detection needs a knowledge-based fact-checking approach, there are similarities between fake news (in its narrow definition) detection methods to other related concepts.
This similarity is due to the fact that fake news still has an information nature and needs a way through social network and conventional news channels to find its audience.

These similarities open the door for leveraging other techniques (e.g., style based and social context based approaches) in detecting fake news content.
However, these solutions are considered incomplete for the problem of fake news if they are not verified by supporting facts and justifications.
Therefore, knowledge-based fact checking approaches (either performed by human or machine learning) become dominant solutions for fake news recognition.
This suggests an extra fact-checking step after the detection step.
The extra recognition step is similar to rumor debunking confirmation in early detection method~\cite{ma2016detecting}.
The necessity of fact-checking step in fake news recognition problem is because fake news detection needs to be verifiable by either supporting facts, justifications, or explanations.
The fake news solution clarification tells us fake news detection without performing the fact checking would present an incomplete solution.

%% file: files/3.2.newsfeed.tex
\subsection{The (Un)accountability of News Feed Algorithms}

Nowadays, personalized news feed algorithms on social media provide content to users based on users' profiles, interests, social media network, and other past click behavior.
Although these algorithms have a key role in the news life cycle, there is not enough research on designing news feed algorithms robust to fake news propagation.

In the following, we review open issues in personalized news feed and search algorithms that have a positive effect on propagation and believe of fake news. 

\subsubsection{Echo Chambers in Social Media}
\label{echo-chamber}
Echo chamber in social media describes a phenomenon where homogeneous views are reinforced by communication inside closed social media groups. 
Echo chambers have been previously studied in relation to creating polarized opinions and shaping a false sense of credibility for users whose frequent news sources are through social media~\cite{zajonc2001mere}.
This false sense of credibility holds users in a vulnerable position of accepting biased and fake news content. % paul2016russian
In similar circumstances, although news feed algorithms are meant to provide content related to users interest, these machine learning algorithms can adjust a user's news feed with a certain perception of reality and trigger user confirmation bias to over-trust unreliable sources~\cite{quattrociocchi2016echo}. 
In studying methods to overcome echo chambers in social media, Lex et al.~\shortcite{lex2018mitigating} presented a content-based news recommendation that can increase exposure of the opposite view to users in order to mitigate the echo chamber effect in social media.
In another work, Hou et al.~\shortcite{hou2018balancing} demonstrated methods to balance popularity bias in network-based recommendation systems and to significantly improve the system's news diversity.

\subsubsection{Filter Bubbles in Search Engines} %  and user isolation}

\textit{Filter bubble} is another term to describe the negative effects of personalized news search engines~\cite{pariser2011filter}.
Filter bubbles represent a state of information isolation where users are only exposed to a certain perspective of information brought to them by personalized search engines.
The lack of exposure to diverse viewpoints creates a filter bubble for individuals and increase the chance of accepting fake content, accrediting unreliable sources, and further distributing fake content.
In the context of news search engines, Haim et al.~\shortcite{haim2018burst} observed negative results in an exploratory analysis of personalized news search engine effect on news diversity.
One approach to encounter unwanted negative effects of these algorithms is to define new measures and standards for sensitive data and products recommendation systems.
For better measuring information diversity in recommender systems, Ekstrand et al.~\shortcite{valcarce2018robustness} and Valcarce et al.~\shortcite{ekstrand2018all} proposed new evaluation measures to account for other considerations like users' popularity bias or content diversity.
The importance of new measures for news recommendation algorithms comes from considering the diversity of content across different user groups.

The importance of news feed algorithms is in their ability to increase information and opinion diversity in social media.
News diversity in social media is another approach to combat fake news by eliminating echo chambers and biased search engines. 

%% file: files/4-discussion.tex
% \Title{Interpretable News Feed Algorithm}
% \section{Interpretable News Recommender Algorithms}
% \section{Interpretable News Feed Algorithms}
\section{Opportunities in Combating Fake News}
\label{Opportunities-section}

Considering the recent attention in designing end-to-end interpretable fake news detection systems (e.g.,~\cite{yang2019XFake} and~\cite{popat2018credeye}), the necessity of interpretability, and variety of social media data modality; it is essential to take new interpretable design approaches to improve model robustness.
In this section, we discuss two research opportunities for fake new mitigation focused on fake news detection methods, news diversity metrics, news feed algorithms, and their evaluation measures.

%% file: files/4.1.human-interpret.tex
\subsection{Interpretable Fake News Detection}
% \subsection{Interpretability in All its Dimensions}

While interpretable machine learning can help model designer for debugging and model validation~\cite{zeiler2014visualizing}, machine learning explanations may also help end-users understand and trust the machine learning systems.
However, appropriate explanation type for machine learning experts is different from novice end-users and media experts.
Figure~\ref{fig:dimensions} shows three dimensions of interpretability that serve different purposes in intelligent fake news mitigation systems.
In the following, we discuss how these three dimensions can improve fake news detection research.

\subsubsection{Algorithmic Interpretability}
% what
Interpretability of algorithms is usually defined as a degree of human understandability of a algorithms decision-making process.
We introduce the \textit{Algorithmic Interpretability} term as a degree of machine learning experts ability to visualize model parameters and inspect model behavior.
% why
Algorithmic interpretability helps in machine learning designers to debug and tune model parameters for more accurate and reliable models.
Machine learning system transparency also leads to evaluate model robustness (for reliability and safety) and model fairness (not being biased). 
% how
Researchers implemented various interpretability methods to explain individual input instances (e.g., feature importance) and visualizing the whole model (e.g., model internal weights).
Fake news detection algorithms can also benefit from algorithmic interpretability and transparency to verify model fairness toward different speakers and topics.

\subsubsection{Human Interpretability}
% Introduction                   
Although beneficial for experts to understand the model behavior, algorithmic interpretability methods are not necessarily useful and even understandable for machine learning product end-users.
We introduce \textit{Human Interpretability} as a product of machine learning interpretability methods that can significantly benefit fake news detection systems.
Human interpretability provides decision-making transparency to end-users with understandable explanations of ``how the system works'' and ``why this decision is made''.
% Why
These explanations may help the user to better understand model prediction details, increase user trust on machine decisions, and therefore increase human-machine performance on the tasks.
Human-computer interaction (HCI) research shows machine learning explanations increase the overall performance of end-users by increasing user knowledge and reliance on the system~\cite{kulesza2015principles}.
% Evaluation
Moreover, human interpretability opens new evaluation methods and measures for the interpretability method.
Different human-subject studies in HCI research evaluate interpretability measures such as user understanding, mental model, explanation usefulness, trust, and task performance~\cite{mohseni2018survey}.

\subsubsection{Supporting Evidence}
%what
In the context of fact-checking and fake news detection, providing supporting evidence is an essential element to justify the decision made by experts and organizations. % fake news identification 
Supporting evidence are accurate information which have relationship to the news event and can verify the veracity of the news.
%why
Providing supporting evidence from the training data or the knowledge graph can provide extensive \textit{why explanation} for each instance. 
%how
Related to this, new research by Yang et al.~\shortcite{yang2019XFake} and~\shortcite{popat2018declare} propose explainable
fake news detection methods that assist end-users to identify news credibility with supporting evidence selected from a set of verified news.
We suggest designing interpretable algorithm that can explain their decisions by providing supporting examples from training data can help users better understand the system and trust the detection results.

%% file: files/4.2.transparency.tex
\begin{figure}[t!]
\centering
  \includegraphics[width=0.7\columnwidth]{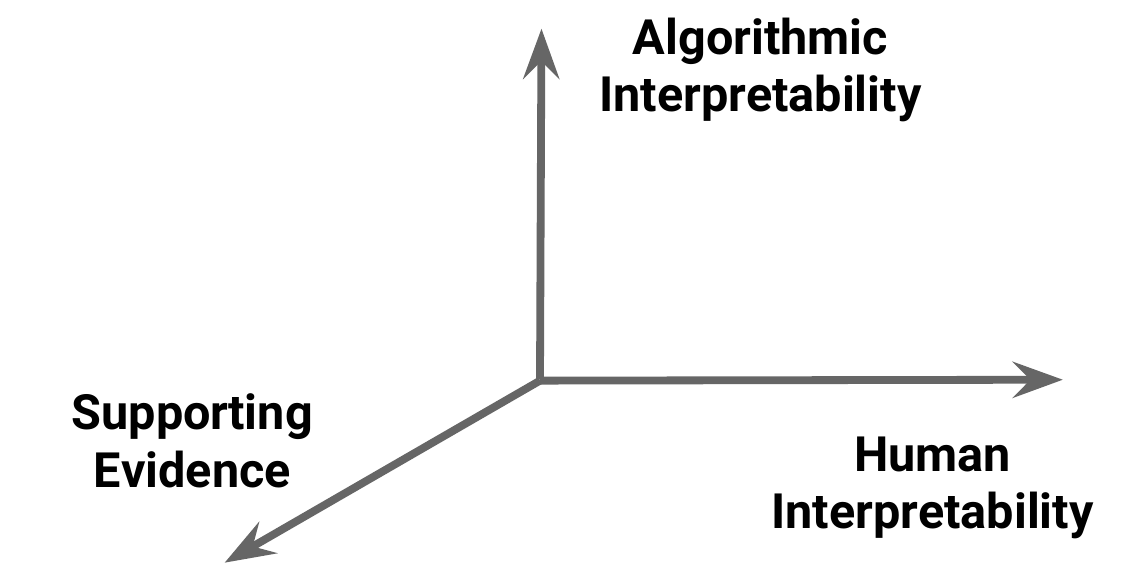}
  \caption{Three dimensions of interpretability in fake news detection systems. Algorithmic interpretability helps machine learning experts to debug and tune machine learning models. Human interpretability provides end-user understandable explanations to trust the machine learning product. Supporting evidence is an inevitable element of news fact-checking that can be from training data or knowledge graphs.} % Each dimension serves different purposes. 
  \label{fig:dimensions}
\end{figure}

% Link: https://docs.google.com/presentation/d/1UQv60Ol09eaHzGHRUUhaAxIsD27JltcuO-o1Xwqvq7M/edit#slide=id.p

% Link new: https://docs.google.com/presentation/d/1SXYMAh6EvewzvuFjqB8zUOTIoUPsXRH_2U5Q3zvs3kI/edit#slide=id.p

\subsection{Interpretable News Feed Algorithms}

News feed algorithms have an important role in content and news distribution in social media (see Figure~\ref{fig:life-cycle}).
In this section, we discuss two different machine learning interpretability approaches that can benefit the mitigation of fake news propagation in social media.

\subsubsection{Interpretability for Robust News Feed}
% \subsubsection{Robust News Feed}

%what
Another fake news mitigation approach is to embed news veracity checking tools in news feed systems to provide explanations about the news content veracity to the user.
The news feed algorithm can perform high-level veracity analysis (e.g., source credibility, deceptive language flag, click bait flag) on social media posts to assist the users with data-driven veracity check results.
Although interpretable news feed is not detection or debunking solution, in real scenarios, mitigation solutions could be more advantageous than automatically debunking of the news. 

%why
The advantages of fake news resistant news feed are in two folds. 
First, veracity checking of content in the distribution step benefits both news consumers and content providers in contrast to auto content removal from the network.
For the user side, providing veracity analysis does not limit the user access to information (in contrast to debunking news) and yet help the users to decide by providing interpretable explanations.
For the content provider side, veracity analysis of misinformation such as satire news and click baits (for entertainment and advertisement purposes) does not automatically remove the content and may still cause damage to the news and content providers.
Second, by looking at news life cycle stages in Figure~\ref{fig:life-cycle}, veracity checking of the news on the distribution step is the faster method compared to other social-context-dependent methods such as rumor detection.

% -- end -- 
We argue that the effectiveness of fake news mitigation algorithms will further extend if being able to provide useful veracity checking explanations for the end user.
Such explanations designed for news feed and search algorithms could benefit both expert journalists and non-expert news consumers.

\subsubsection{Interpretability for User Awareness}
% Combating
Algorithmic transparency is another way that social media can benefit from interpretable machine learning to mitigate the distribution of fake content. 
News feed algorithms in social media process sensitive user data to generate a user's reading list and do not give options for the user to choose among content.
In these circumstances, the lack of explanation of ``how the content is selected?'' for the user may bias the user toward what the algorithm is providing in the news feed and not the entire reality. 

While transparency is not a silver bullet to stop fake news propagation in social media, it helps the user to understand how news feed algorithms work to use them appropriately.
For example, ``model visualization of user preference'' (e.g., in form of user model) and ``news attributes that contribute to the news recommendation'' (e.g., features weights ) will help users to understand how the content selection algorithm is working.
Other explanations such as ``what kind of \textit{soft filters} are applied'' on users feed also would benefit on the purpose of user awareness.
News feed transparency through model explanation could also reduce the occurrence of large scale adverse situations such as propagation of biased and fake news in social media.

We suggest adding model and instance explanations to news feed algorithms would serve as a major leap toward accountable news feed systems.
Such transparency could defeat fake new by increasing user awareness and eliminating situations such as echo chambers and filter bubbles. % the propagation of fake new 

%% file: files/5-conclusion.tex
\section{Conclusion}
We reviewed the opportunities and limitations in the current research on fake news detection methods and the transparency of news feed algorithms.
After reviewing current fake news detection methods along three main stages of news life in social media,
we discuss two open issues in machine learning fake news research.
In the first discussion, we clarified the issues in fake news detection problem characterization, 
Then we emphasized the vulnerability of news feed algorithms in propagating fake news.
In the last section, we presented interpretability with its three dimensions as a potential solution and direction for fake news research that can benefit both machine learning experts and social media end-users.
We also introduced interpretable news feed systems as an effective fake news elimination solution in real scenarios compared to the current detection and debunking solutions.

% \section{Acknowledgments}
% This is a placeholder for the acknowledgments section.

%% file: mian.bbl
\begin{thebibliography}{}

\bibitem[\protect\citeauthoryear{Afchar \bgroup \em et al.\egroup
  }{2018}]{afchar2018mesonet}
Darius Afchar, Vincent Nozick, Junichi Yamagishi, and Isao Echizen.
\newblock Mesonet: a compact facial video forgery detection network.
\newblock In {\em 2018 IEEE International Workshop on Information Forensics and
  Security (WIFS)}, pages 1--7. IEEE, 2018.

\bibitem[\protect\citeauthoryear{Afroz \bgroup \em et al.\egroup
  }{2012}]{afroz2012detecting}
Sadia Afroz, Michael Brennan, and Rachel Greenstadt.
\newblock Detecting hoaxes, frauds, and deception in writing style online.
\newblock In {\em Security and Privacy (SP), 2012 IEEE Symposium on}, pages
  461--475. IEEE, 2012.

\bibitem[\protect\citeauthoryear{Allcott and
  Gentzkow}{2017}]{allcott2017social}
Hunt Allcott and Matthew Gentzkow.
\newblock Social media and fake news in the 2016 election.
\newblock {\em Journal of Economic Perspectives}, 31(2):211--36, 2017.

\bibitem[\protect\citeauthoryear{Bessi and Ferrara}{2016}]{bessi2016social}
Alessandro Bessi and Emilio Ferrara.
\newblock Social bots distort the 2016 us presidential election online
  discussion.
\newblock 2016.

\bibitem[\protect\citeauthoryear{Bharati \bgroup \em et al.\egroup
  }{2018}]{bharati2018beyond}
Aparna Bharati, Daniel Moreira, Joel Brogan, Patricia Hale, Kevin~W Bowyer,
  Patrick~J Flynn, Anderson Rocha, and Walter~J Scheirer.
\newblock Beyond pixels: Image provenance analysis leveraging metadata.
\newblock {\em arXiv preprint arXiv:1807.03376}, 2018.

\bibitem[\protect\citeauthoryear{Bozdag}{2013}]{bozdag2013bias}
Engin Bozdag.
\newblock Bias in algorithmic filtering and personalization.
\newblock {\em Ethics and information technology}, 15(3):209--227, 2013.

\bibitem[\protect\citeauthoryear{Castillo \bgroup \em et al.\egroup
  }{2013}]{castillo2013predicting}
Carlos Castillo, Marcelo Mendoza, and Barbara Poblete.
\newblock Predicting information credibility in time-sensitive social media.
\newblock {\em Internet Research}, 23(5):560--588, 2013.

\bibitem[\protect\citeauthoryear{Ekstrand \bgroup \em et al.\egroup
  }{2018}]{ekstrand2018all}
Michael~D Ekstrand, Mucun Tian, Ion~Madrazo Azpiazu, Jennifer~D Ekstrand,
  Oghenemaro Anuyah, David McNeill, and Maria~Soledad Pera.
\newblock All the cool kids, how do they fit in?: Popularity and demographic
  biases in recommender evaluation and effectiveness.
\newblock In {\em Conference on Fairness, Accountability and Transparency},
  pages 172--186, 2018.

\bibitem[\protect\citeauthoryear{Geschke \bgroup \em et al.\egroup
  }{2018}]{geschke2018triple}
Daniel Geschke, Jan Lorenz, and Peter Holtz.
\newblock The triple-filter bubble: Using agent-based modelling to test a
  meta-theoretical framework for the emergence of filter bubbles and echo
  chambers.
\newblock {\em British Journal of Social Psychology}, 2018.

\bibitem[\protect\citeauthoryear{Gunning}{2017}]{gunning2017explainable}
David Gunning.
\newblock Explainable artificial intelligence (xai).
\newblock {\em Defense Advanced Research Projects Agency (DARPA), nd Web},
  2017.

\bibitem[\protect\citeauthoryear{Haim \bgroup \em et al.\egroup
  }{2018}]{haim2018burst}
Mario Haim, Andreas Graefe, and Hans-Bernd Brosius.
\newblock Burst of the filter bubble? effects of personalization on the
  diversity of google news.
\newblock {\em Digital Journalism}, 6(3):330--343, 2018.

\bibitem[\protect\citeauthoryear{Hou \bgroup \em et al.\egroup
  }{2018}]{hou2018balancing}
Lei Hou, Xue Pan, and Kecheng Liu.
\newblock Balancing the popularity bias of object similarities for personalised
  recommendation.
\newblock {\em The European Physical Journal B}, 91(3):47, 2018.

\bibitem[\protect\citeauthoryear{Jin \bgroup \em et al.\egroup
  }{2016}]{jin2016news}
Zhiwei Jin, Juan Cao, Yongdong Zhang, and Jiebo Luo.
\newblock News verification by exploiting conflicting social viewpoints in
  microblogs.
\newblock In {\em AAAI}, pages 2972--2978, 2016.

\bibitem[\protect\citeauthoryear{Julliand \bgroup \em et al.\egroup
  }{2015}]{julliand2015image}
Thibaut Julliand, Vincent Nozick, and Hugues Talbot.
\newblock Image noise and digital image forensics.
\newblock In {\em International Workshop on Digital Watermarking}, pages 3--17.
  Springer, 2015.

\bibitem[\protect\citeauthoryear{Kalogeropoulos and
  Nielsen}{2018}]{Social2018Inequalities}
Antonis Kalogeropoulos and Rasmus~Kleis Nielsen.
\newblock Social inequalities in news consumption.
\newblock In {\em FACTSHEET, NEWS MEDIA DIGITAL MEDIA}, pages 461--475. Reuters
  Institute for the Study of Journalism, 2018.

\bibitem[\protect\citeauthoryear{Kulesza \bgroup \em et al.\egroup
  }{2015}]{kulesza2015principles}
Todd Kulesza, Margaret Burnett, Weng-Keen Wong, and Simone Stumpf.
\newblock Principles of explanatory debugging to personalize interactive
  machine learning.
\newblock In {\em Proceedings of the 20th international conference on
  intelligent user interfaces}, pages 126--137. ACM, 2015.

\bibitem[\protect\citeauthoryear{Kwon \bgroup \em et al.\egroup
  }{2013}]{kwon2013prominent}
Sejeong Kwon, Meeyoung Cha, Kyomin Jung, Wei Chen, and Yajun Wang.
\newblock Prominent features of rumor propagation in online social media.
\newblock In {\em 2013 IEEE 13th International Conference on Data Mining},
  pages 1103--1108. IEEE, 2013.

\bibitem[\protect\citeauthoryear{Lex \bgroup \em et al.\egroup
  }{2018}]{lex2018mitigating}
Elisabeth Lex, Mario Wagner, and Dominik Kowald.
\newblock Mitigating confirmation bias on twitter by recommending opposing
  views.
\newblock {\em arXiv preprint arXiv:1809.03901}, 2018.

\bibitem[\protect\citeauthoryear{Ma \bgroup \em et al.\egroup
  }{2016}]{ma2016detecting}
Jing Ma, Wei Gao, Prasenjit Mitra, Sejeong Kwon, Bernard~J Jansen, Kam-Fai
  Wong, and Meeyoung Cha.
\newblock Detecting rumors from microblogs with recurrent neural networks.
\newblock In {\em IJCAI}, pages 3818--3824, 2016.

\bibitem[\protect\citeauthoryear{Mohseni \bgroup \em et al.\egroup
  }{2018}]{mohseni2018survey}
Sina Mohseni, Niloofar Zarei, and Eric~D Ragan.
\newblock A survey of evaluation methods and measures for interpretable machine
  learning.
\newblock {\em arXiv preprint arXiv:1811.11839}, 2018.

\bibitem[\protect\citeauthoryear{Pariser}{2011}]{pariser2011filter}
Eli Pariser.
\newblock {\em The filter bubble: What the Internet is hiding from you}.
\newblock Penguin UK, 2011.

\bibitem[\protect\citeauthoryear{Popat \bgroup \em et al.\egroup
  }{2018a}]{popat2018credeye}
Kashyap Popat, Subhabrata Mukherjee, Jannik Str{\"o}tgen, and Gerhard Weikum.
\newblock Credeye: A credibility lens for analyzing and explaining
  misinformation.
\newblock In {\em Companion of the The Web Conference 2018 on The Web
  Conference 2018}, pages 155--158. International World Wide Web Conferences
  Steering Committee, 2018.

\bibitem[\protect\citeauthoryear{Popat \bgroup \em et al.\egroup
  }{2018b}]{popat2018declare}
Kashyap Popat, Subhabrata Mukherjee, Andrew Yates, and Gerhard Weikum.
\newblock Declare: Debunking fake news and false claims using evidence-aware
  deep learning.
\newblock {\em arXiv preprint arXiv:1809.06416}, 2018.

\bibitem[\protect\citeauthoryear{Potthast \bgroup \em et al.\egroup
  }{2016}]{potthast2016clickbait}
Martin Potthast, Sebastian K{\"o}psel, Benno Stein, and Matthias Hagen.
\newblock Clickbait detection.
\newblock In {\em European Conference on Information Retrieval}, pages
  810--817. Springer, 2016.

\bibitem[\protect\citeauthoryear{Potthast \bgroup \em et al.\egroup
  }{2017}]{potthast2017stylometric}
Martin Potthast, Johannes Kiesel, Kevin Reinartz, Janek Bevendorff, and Benno
  Stein.
\newblock A stylometric inquiry into hyperpartisan and fake news.
\newblock {\em arXiv preprint arXiv:1702.05638}, 2017.

\bibitem[\protect\citeauthoryear{Quattrociocchi \bgroup \em et al.\egroup
  }{2016}]{quattrociocchi2016echo}
Walter Quattrociocchi, Antonio Scala, and Cass~R Sunstein.
\newblock Echo chambers on facebook.
\newblock 2016.

\bibitem[\protect\citeauthoryear{Rubin and Lukoianova}{2015}]{rubin2015truth}
Victoria~L Rubin and Tatiana Lukoianova.
\newblock Truth and deception at the rhetorical structure level.
\newblock {\em Journal of the Association for Information Science and
  Technology}, 66(5):905--917, 2015.

\bibitem[\protect\citeauthoryear{Rubin \bgroup \em et al.\egroup
  }{2016}]{rubin2016fake}
Victoria Rubin, Niall Conroy, Yimin Chen, and Sarah Cornwell.
\newblock Fake news or truth? using satirical cues to detect potentially
  misleading news.
\newblock In {\em Proceedings of the Second Workshop on Computational
  Approaches to Deception Detection}, pages 7--17, 2016.

\bibitem[\protect\citeauthoryear{Sharma \bgroup \em et al.\egroup
  }{2019}]{sharma2019combating}
Karishma Sharma, Feng Qian, He~Jiang, Natali Ruchansky, Ming Zhang, and Yan
  Liu.
\newblock Combating fake news: A survey on identification and mitigation
  techniques.
\newblock {\em arXiv preprint arXiv:1901.06437}, 2019.

\bibitem[\protect\citeauthoryear{Shu \bgroup \em et al.\egroup
  }{2017a}]{shu2017fake}
Kai Shu, Amy Sliva, Suhang Wang, Jiliang Tang, and Huan Liu.
\newblock Fake news detection on social media: A data mining perspective.
\newblock {\em ACM SIGKDD Explorations Newsletter}, 19(1):22--36, 2017.

\bibitem[\protect\citeauthoryear{Shu \bgroup \em et al.\egroup
  }{2017b}]{shu2017exploiting}
Kai Shu, Suhang Wang, and Huan Liu.
\newblock Exploiting tri-relationship for fake news detection.
\newblock {\em arXiv preprint arXiv:1712.07709}, 2017.

\bibitem[\protect\citeauthoryear{Shu \bgroup \em et al.\egroup
  }{2018}]{shu2018fakenewsnet}
Kai Shu, Deepak Mahudeswaran, Suhang Wang, Dongwon Lee, and Huan Liu.
\newblock Fakenewsnet: A data repository with news content, social context and
  dynamic information for studying fake news on social media.
\newblock {\em arXiv preprint arXiv:1809.01286}, 2018.

\bibitem[\protect\citeauthoryear{Spirin and Han}{2012}]{spirin2012survey}
Nikita Spirin and Jiawei Han.
\newblock Survey on web spam detection: principles and algorithms.
\newblock {\em ACM SIGKDD explorations newsletter}, 13(2):50--64, 2012.

\bibitem[\protect\citeauthoryear{Tacchini \bgroup \em et al.\egroup
  }{2017}]{tacchini2017some}
Eugenio Tacchini, Gabriele Ballarin, Marco~L Della~Vedova, Stefano Moret, and
  Luca de~Alfaro.
\newblock Some like it hoax: Automated fake news detection in social networks.
\newblock {\em arXiv preprint arXiv:1704.07506}, 2017.

\bibitem[\protect\citeauthoryear{Valcarce \bgroup \em et al.\egroup
  }{2018}]{valcarce2018robustness}
Daniel Valcarce, Alejandro Bellog{\'\i}n, Javier Parapar, and Pablo Castells.
\newblock On the robustness and discriminative power of information retrieval
  metrics for top-n recommendation.
\newblock In {\em Proceedings of the 12th ACM Conference on Recommender
  Systems}, pages 260--268. ACM, 2018.

\bibitem[\protect\citeauthoryear{Vlachos and Riedel}{2014}]{vlachos2014fact}
Andreas Vlachos and Sebastian Riedel.
\newblock Fact checking: Task definition and dataset construction.
\newblock In {\em Proceedings of the ACL 2014 Workshop on Language Technologies
  and Computational Social Science}, pages 18--22, 2014.

\bibitem[\protect\citeauthoryear{Yang \bgroup \em et al.\egroup
  }{2019}]{yang2019XFake}
Fan Yang, Shiva~K. Pentyala, Sina Mohseni, Mengnan Du, Hao Yuan, Rhema Linder,
  Eric~D. Ragan, Shuiwang Ji, and Xia~(Ben) Hu.
\newblock Xfake: Explainable fake news detector with visualizations.
\newblock In {\em Companion of the The Web Conference 2019 on The Web
  Conference 2018}, pages 155--158. ACM, 2019.

\bibitem[\protect\citeauthoryear{Zajonc}{2001}]{zajonc2001mere}
Robert~B Zajonc.
\newblock Mere exposure: A gateway to the subliminal.
\newblock {\em Current directions in psychological science}, 10(6):224--228,
  2001.

\bibitem[\protect\citeauthoryear{Zeiler and
  Fergus}{2014}]{zeiler2014visualizing}
Matthew~D Zeiler and Rob Fergus.
\newblock Visualizing and understanding convolutional networks.
\newblock In {\em European conference on computer vision}, pages 818--833.
  Springer, 2014.

\bibitem[\protect\citeauthoryear{Zhou and Zafarani}{2018}]{zhou2018fake}
Xinyi Zhou and Reza Zafarani.
\newblock Fake news: A survey of research, detection methods, and
  opportunities.
\newblock {\em arXiv preprint arXiv:1812.00315}, 2018.

\bibitem[\protect\citeauthoryear{Zubiaga \bgroup \em et al.\egroup
  }{2018}]{zubiaga2018detection}
Arkaitz Zubiaga, Ahmet Aker, Kalina Bontcheva, Maria Liakata, and Rob Procter.
\newblock Detection and resolution of rumours in social media: A survey.
\newblock {\em ACM Computing Surveys (CSUR)}, 51(2):32, 2018.

\end{thebibliography}
